\documentclass{emulateapj}

\usepackage{multirow}

\begin{document}

\title{Post-Capture Evolution of Potentially Habitable Exomoons}
\author{Simon B. Porter}
\email{porter@lowell.edu}
\affil{Lowell Observatory, 1400 W. Mars Hill Rd., Flagstaff AZ 86001}
\affil{School of Earth and Space Exploration, Arizona State University, Tempe AZ 85287}

\and

\author{William M. Grundy}
\affil{Lowell Observatory, 1400 W. Mars Hill Rd., Flagstaff AZ 86001}

\begin{abstract}
The satellites of extrasolar planets (exomoons) have been recently proposed as astrobiological targets. 
Since giant planets in the habitable zone are thought to have migrated there, it is possible that they may have captured a former terrestrial planet or planetesimal. 
We therefore attempt to model the dynamical evolution of a terrestrial planet captured into orbit around a giant planet in the habitable zone of a star. 
We find that approximately half of loose elliptical orbits result in stable circular orbits over timescales of less than a few million years. 
We also find that those orbits are mostly low-inclination, but have no prograde/retrograde preference.
In addition, we calculate the transit timing and duration variations for the resulting systems, and find that potentially habitable Earth-mass exomoons should be detectable.
\end{abstract}


\maketitle

\section{Motivation}

Exomoons, the satellites of extrasolar planets, have been often featured in fiction as habitable locations. 
There is no deficit of known giant planets; Exoplanet.org \citep{ExoplanetOrg} lists approximately 40 giant exoplanets (8\% of total) within 20\% of the equilibrium temperature of Earth, 
as are 30 (3\%) of the \textit{Kepler} planet candidates released in February 2011 \citep{Borucki2011}.
Though these observations are preliminary, they do show that habitable-zone giant planets not only exist, but are common.
Once a giant planet is known to be in a habitable zone, variations in its orbit, such as Transit Timing Variation \citep[TTV;][]{Sartoretti1999} and Transit Duration Variation \citep[TDV;][]{Kipping2009},
photometry \citep{Szabo2006}, or gravitational microlensing \citep{Liebig2010}, allow the indirect detection of satellites.
Thus, if potentially habitable exomoons exist around transiting giant planets, they may be detected at the same (or even greater) rate as solitary habitable terrestrial planets.
As yet, no exomoons have been detected, but the wealth of transit data from the Kepler mission should begin to fill this gap.

Despite the existence of giant planets in stellar habitable zones, it is far from certain how they arrived there. 
Current giant-planet formation models assume that they are created at distances beyond the stability point of ice \citep[e.g.][]{Lissauer1987,Boss1997}, which implies conditions not suitable to surface habitability.
Disk migration can bring giant planets close to the star \citep{Ward1997}, but generally has a stopping point far too close to the star to be habitable (thus producing "Hot Jupiters").
The host planets of potentially habitable exomoons therefore likely arrived at their final orbit through late-stage migration, driven either by planetesimals \citep{Kirsh2009} or other giant planets \citep{Weidenschilling1996}.

In the process of migrating, the satellite systems of these giant planet may have close encounters with terrestrial planets or planetesimals, causing them to be disrupted or replaced.
If either the Jovian or Saturnian systems were transported to 1 AU around a solar mass-star, both Callisto and Titan would be at 18\% of their planet's Hill radii, 
thus implying that all the major satellites of the two planets would be on stable orbits.
However, a close encounter could either excite their orbits to high eccentricity (thus requiring tidal recircularization), or could result in the capture of a much larger terrestrial satellite.
Neptune appears have to experienced this process during its migration through the proto-Kuiper Belt, loosing any original major satellites, while gaining Triton in an inclined, retrograde orbit.
This was possibly due to a momentum-exchange reaction that ejected the binary companion of Triton \citep{Agnor2006}, though other scenarios are possible (at reduced probability).
Any capture process, though, will tend to produce very loosely-bound initial orbits, with only a small delta-v to escape velocity at periapse.
Therefore, some method must be used to determine the long-term evolution and stability (or lack thereof) for these orbits.
Here we use a full KCTF (Kozai Cycle and Tidal Friction) model to find the survival probability for a range of physical conditions and the detectability of the resulting system.

As shown by \citet{Donnison2010} and \citet{Sato2010}, there are a range of stable orbits for Earth-mass planets around giant planets. 
However, both of those models test only the stability of the orbit, rather than any evolution due to tidal effects.
On inclined exomoon orbits, the effects of stellar torques on the orbit can, through initiating Kozai cycles, dramatically accelerate the rate of tidal decay, orbit circularization, and spin-orbit synchronization.
As we will show, this process can allow even very loose, inclined capture orbits to stabilize to tight, circular orbits.
Thus, even marginal dynamical captures of former terrestrial planets can produce stable exomoons.

\begin{figure}
\plotone{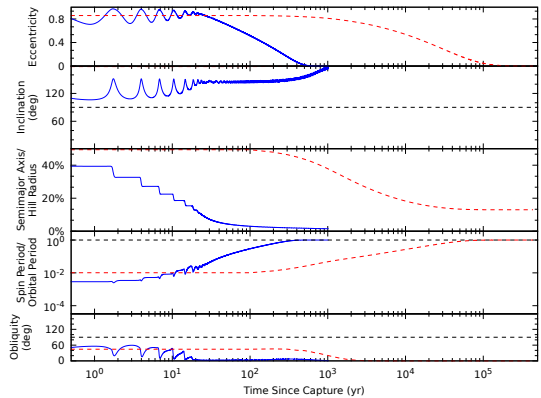}
\caption{Post-capture spin-orbit evolution of two exomoons; the solid line shows a Kozai-enhanced decay, while the dashed line shows a non-Kozai decay.}
\label{fig:tale}
\end{figure}

\section{KCTF Model}

In order to understand how exomoon orbits may evolve after being captured, we created a numerical Kozai Cycle and Tidal Friction (KCTF) model. 
Kozai cycles in this context are the secular oscillations in eccentricity and inclination of the exomoon's orbit caused by stellar torques \citep{Kozai1962}. 
In isolation, these oscillations preserve the semimajor axis of the orbit and the quantity $H_k=cos(I)\times\sqrt{1-e^2}$, 
where $I$ is the inclination of the exomoon's orbit relative to the planet's stellarcentric orbit, and $e$ is the eccentricity of the exomoon's orbit.
In the scenarios simulated, these cycles can initially be as fast as just a few years (see Figure \ref{fig:tale}), causing very rapid evolution of the exomoon's orbit.
Since all the initial orbits for this study were highly elliptical, only very low initial inclinations (within about $15^{\circ}$ of coplanar) produced values of $H_k$ sufficiently low that Kozai cycles did not occur.

Because the Kozai oscillations from the star attempt to preserve $H_k$, a drop in the inclination of the orbit can cause the eccentricity to become very high.
As the eccentricity of the orbit increases, the periapse of the exomoon's orbit becomes much closer to the planet.
Tidal forces become much stronger at these close distances (going as $a^{-8}$ in this model), and so a close periapse due to eccentricity causes the orbit to shrink and thus decay further.
This sets up a positive feedback loop, which progressively shrinks and circularizes the orbit.
The stair-step semimajor axis decay in Figure \ref{fig:tale} happens because the obit is initially only decaying at high eccentricities; 
once the apoapse is close enough to also experience tidal forces, the oscillations stop.
The result is to cause semimajor axis decay and circularization much faster than if the star were not exciting eccentricity.
In addition, since the high eccentricity part of the Kozai cycles are at low inclinations, orbits are preferentially frozen near the plane of the stellarcentric orbit.

To simulate this process, we used a KCTF model based on that of \citet{Eggleton2001}, which allows the direct integration of the exomoon orbit's specific angular momentum vector $\mathbf{h}$ and 
Laplace-Runge-Lenz eccentricity vector $\mathbf{e}$, as well as the spin vectors of the planet and satellite.
The tidal properties of the giant planets were based on those presented in \citet{Fabrycky2007} for a Jupiter-mass planet.
The tidal properties of the exomoons used the formulation of \citet{Ragozzine2009}, along with his addition of solid-body quadrapole gravity.
To integrate the system of equations, we used the Burlisch-Stoer method with vector-rational interpolation \citep{Sweatman1998} 
and error control based on the algorithm given in \citet{Press2007}, with a per-timestep tolerance of $10^{-10}$.

This model does not include any dynamical effects from objects external to the exomoon-exoplanet system other than the star itself.
In addition, we assumed that the planet did not migrate over the course of the simulation.
If it were to migrate, the effect would be to shrink the Hill radius of the planet and shorten the period of the Kozai cycles.
The smaller Hill radii would allow more captured satellites to escape, while on the other hand, the faster Kozai cycles would decrease the decay timescale.
Thus, if this model works fast enough for a static case, it should also be applicable to a slowly migrating planet.

\section{Monte Carlo Simulations}

To find in what conditions a captured exomoon may survive, we generated 18 sets of synthetic Star-Planet-Moon systems and performed KCTF simulations on them.
Each set contained 1000 synthetic systems, with common masses for all objects, and randomized initial exomoon orbits and spin states.
To simulate a loose capture, the initial exomoon orbits all had apoapses beyond 80\% of the planet's Hill radius, and eccentricities greater than 0.85.
This is most consistent with a low mass ratio momentum-exchange capture \citep{Funato2004}, but generally similar to any low delta-v, non-disruptive capture (or eccentricity excitation due to a close encounter).
Again consistent with a capture, both the satellite's initial orbital plane and spin vector were initially pointed at random directions on the sky.
This means an approximate equipartition between prograde and retrograde initial orbits and between prograde and retrograde initial spin states for the satellite.
The planets had a random obliquity less than $5^{\circ}$ from their stellarcentric orbit.
Each planet-moon system was at a stellarcentric distance such that the equilibrium temperature was equal to Earth.
The stars and stellarcentric orbits used were the Sun (G2) at 1.0 AU, a main-sequence F0 (1.7 $M_{Sun}$) at 2.1 AU, and a main-sequence M0 (0.47 $M_{Sun}$) at 0.28 AU.
The planets we assumed to have a mass equal to either Jupiter or Neptune, using the tidal parameters given in \citet{Fabrycky2007}, though nearly all dissipation was in the exomoon.
The simulated exomoons had the mass of either Earth, Mars, or Titan (with Mars uncompressed density), using a tidal $Q$ of $100$, modulus of rigidity of $3\times10^{10} N/m^2$ \citep{Gladman1996}, and $J_2$ of $0.001$.
The initial rotational periods of the planet and satellite were varied randomly between 0.1 and 48 hours.
The simulations were run until they reached either $10^9$ years or an eccentricity below $10^{-5}$.
However, the simulations were terminated early if the periapse went below the Roche limit (thus potentially causing breakup of the exomoon) or the apoapse exceeded the Hill radius (allowing the exomoon to become unbound).
It is possible that the exomoon could have survived and remained bound to the planet in these scenarios, but that is beyond the fidelity of these simulations.

To estimate the detectability of the resulting systems, we calculated root-mean-squared Transit Timing Variations (TTVs) and Duration Variations (TDVs) for each simulation.
The TTV and TDV are due to the orbit of the exomoon causing the exoplanet to begin the transit either sooner or later than the barycenter of the planet-moon system.
These effects are maximized for low mass ratios (e.g. Earth mass exomoon around a Neptune-mass planet) and longer period exomoon orbits.
Since the majority of resulting systems were low-inclination with respect to the stellarcentric orbit, we used the zero-inclination equations from \citet{Kipping2009}.
Assuming zero-inclination maximizes the TTV and TDV, allowing an estimate of whether the system would be even detectable in the best case scenario.

\begin{figure}
\plottwo{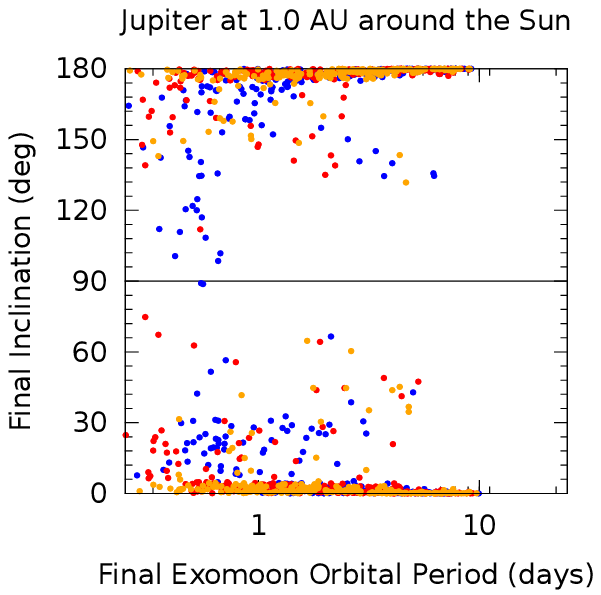}{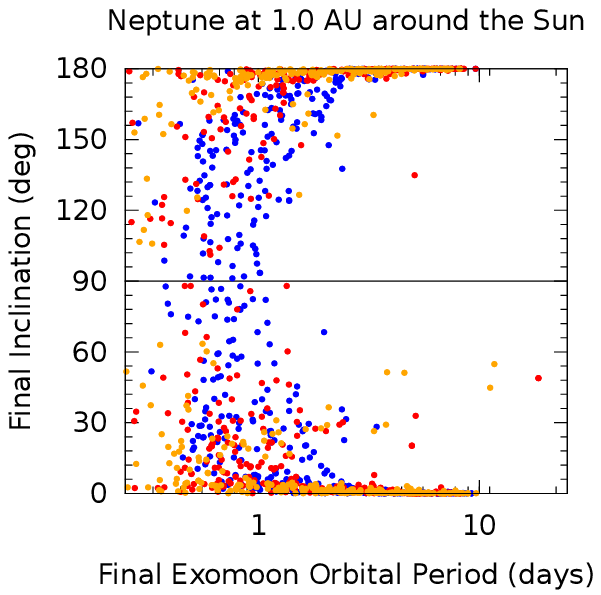} \\
\plottwo{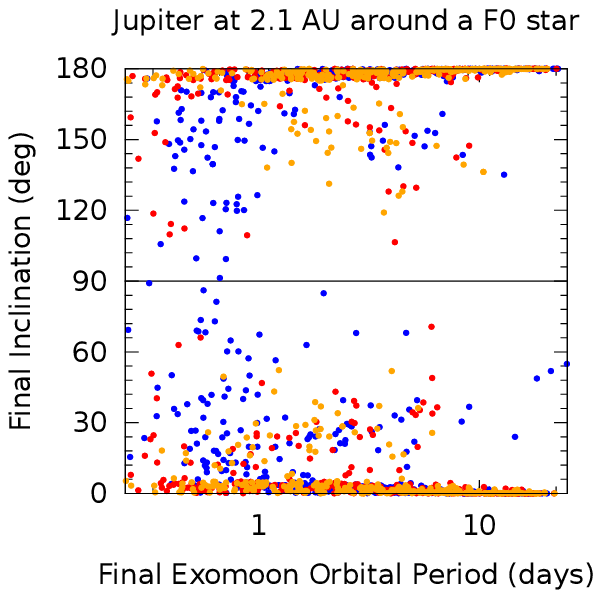}{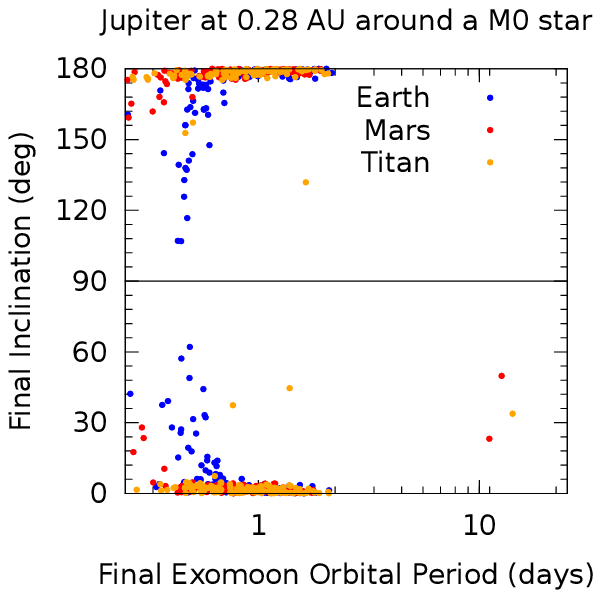}
\caption{Post-evolution orbital period and inclination distribution; most orbits are coplanar with the planet's orbit, with no pro/retrograde preference.}
\label{fig:inc}
\end{figure}

\begin{figure}
\plottwo{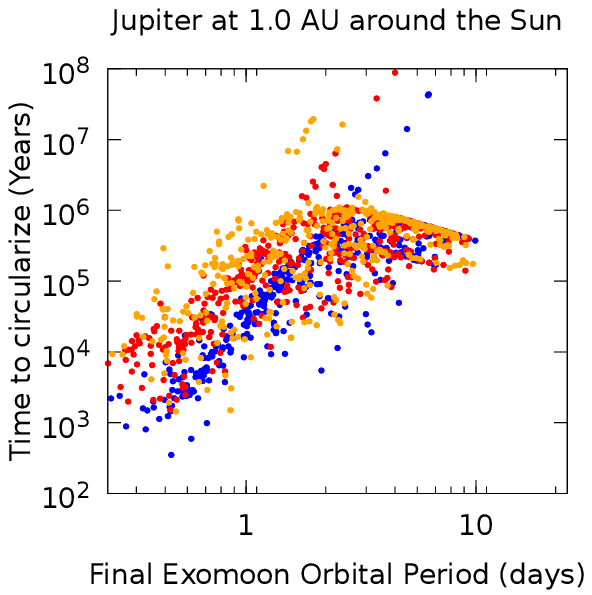}{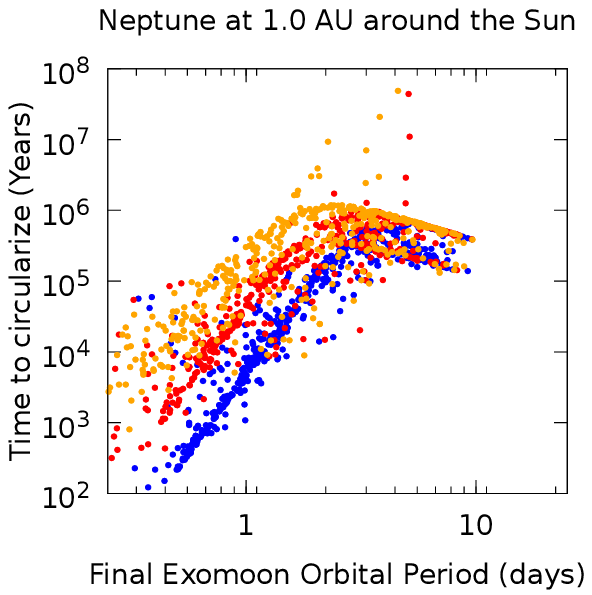} \\
\plottwo{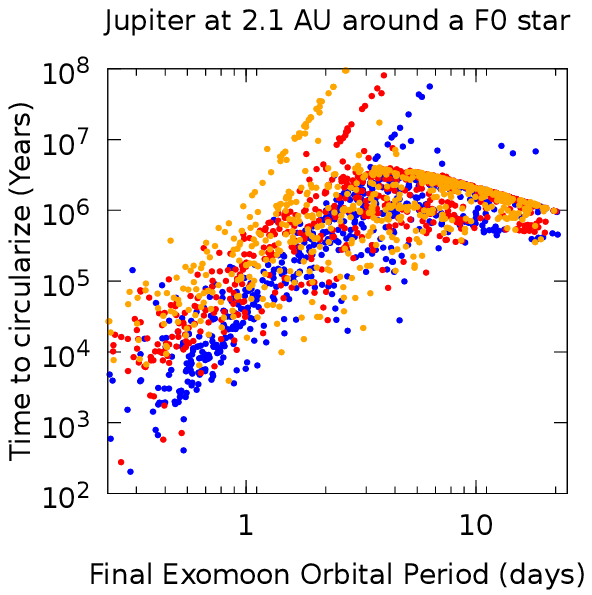}{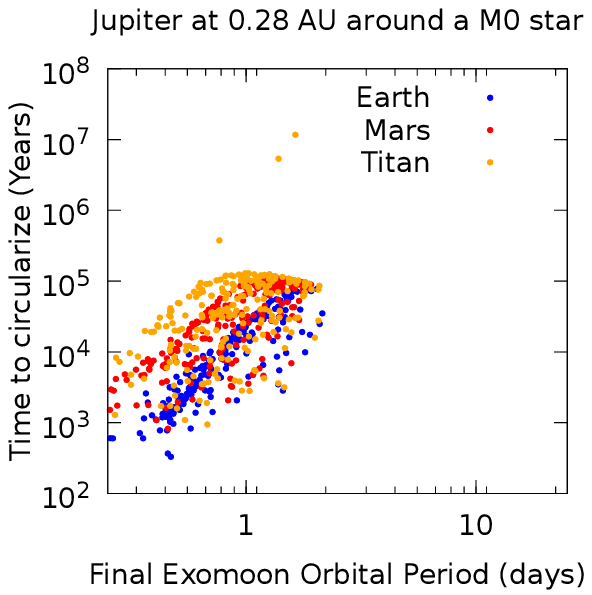}
\caption{Post-capture circularization timescales; note that nearly all are below a few million years, and timescales decrease with stellar mass.}
\label{fig:circ}
\end{figure}

\begin{deluxetable*}{lllcccc}
\tabletypesize{\scriptsize}
\tablecaption{Relative fraction of end states for fully evolved exomoon systems.\label{tab:gen}}
\tablewidth{0pt} \tablenum{1}
\tablehead{\colhead{Star} & \colhead{Planet} & \colhead{Moon} & \colhead{Survived} & \colhead{Retrograde} & \colhead{Separated} & \colhead{Impacted}}
\startdata
\multirow{6}{*}{Sun} & \multirow{3}{*}{Jupiter} & Earth & 43\% & 52\% & 21\% & 35\% \\
 &  & Mars & 44\% & 45\% & 18\% & 37\% \\
 &  & Titan & 42\% & 47\% & 21\% & 36\% \\
\cline{2-7}
 & \multirow{3}{*}{Neptune} & Earth & 52\% & 44\% & 17\% & 30\% \\
 &  & Mars & 44\% & 45\% & 18\% & 36\% \\
 &  & Titan & 45\% & 47\% & 19\% & 35\% \\
\tableline
\multirow{6}{*}{F0} & \multirow{3}{*}{Jupiter} & Earth & 65\% & 47\% & 3\% & 31\% \\
 &  & Mars & 59\% & 46\% & 4\% & 35\% \\
 &  & Titan & 61\% & 48\% & 3\% & 34\% \\
\cline{2-7}
 & \multirow{3}{*}{Neptune} & Earth & 77\% & 44\% & 4\% & 18\% \\
 &  & Mars & 67\% & 44\% & 4\% & 28\% \\
 &  & Titan & 61\% & 50\% & 4\% & 33\% \\
\tableline
\multirow{6}{*}{M0} & \multirow{3}{*}{Jupiter} & Earth & 23\% & 50\% & 2\% & 74\% \\
 &  & Mars & 23\% & 51\% & 3\% & 73\% \\
 &  & Titan & 23\% & 44\% & 2\% & 74\% \\
\cline{2-7}
 & \multirow{3}{*}{Neptune} & Earth & 24\% & 50\% & 1\% & 73\% \\
 &  & Mars & 25\% & 52\% & 1\% & 73\% \\
 &  & Titan & 22\% & 50\% & 3\% & 74\% \\
\enddata
\end{deluxetable*}

\begin{deluxetable*}{lllcccc}
\tabletypesize{\scriptsize}
\tablecaption{Median Exomoon orbital period, Transit Timing Variation (TTV), and Transit Duration Variation (TDV) for full-evolved exomoon systems.\label{tab:trans}}
\tablewidth{0pt} \tablenum{2}
\tablehead{\colhead{Star} & \colhead{Planet} & \colhead{Moon} & \colhead{Period (d)} & \colhead{TTV (min)} & \colhead{TDV} & \colhead{edge-on (min)}}
\startdata
\multirow{6}{*}{Sun} & \multirow{3}{*}{Jupiter} & Earth & 2.17 & 5.44 & 0.114\% & 0.93 \\
 &  & Mars & 2.34 & 0.59 & 0.012\% & 0.10 \\
 &  & Titan & 2.52 & 0.12 & 0.002\% & 0.02 \\
\cline{2-7}
 & \multirow{3}{*}{Neptune} & Earth & 1.65 & 36.69 & 0.835\% & 6.41 \\
 &  & Mars & 1.89 & 4.11 & 0.089\% & 0.69 \\
 &  & Titan & 2.16 & 0.86 & 0.018\% & 0.14 \\
\tableline
\multirow{6}{*}{F0} & \multirow{3}{*}{Jupiter} & Earth & 3.68 & 11.33 & 0.111\% & 1.45 \\
 &  & Mars & 4.26 & 1.22 & 0.011\% & 0.15 \\
 &  & Titan & 4.14 & 0.26 & 0.002\% & 0.03 \\
\cline{2-7}
 & \multirow{3}{*}{Neptune} & Earth & 2.38 & 75.70 & 0.860\% & 10.60 \\
 &  & Mars & 3.79 & 8.51 & 0.083\% & 1.03 \\
 &  & Titan & 3.50 & 1.78 & 0.018\% & 0.22 \\
\tableline
\multirow{6}{*}{M0} & \multirow{3}{*}{Jupiter} & Earth & 0.94 & 1.63 & 0.122\% & 0.54 \\
 &  & Mars & 1.00 & 0.18 & 0.013\% & 0.06 \\
 &  & Titan & 1.00 & 0.04 & 0.003\% & 0.01 \\
\cline{2-7}
 & \multirow{3}{*}{Neptune} & Earth & 0.85 & 10.97 & 0.836\% & 3.36 \\
 &  & Mars & 0.79 & 1.22 & 0.098\% & 0.39 \\
 &  & Titan & 0.83 & 0.26 & 0.020\% & 0.08 \\
\enddata
\end{deluxetable*}

\section{Results and Discussion}

The general result of the simulations is that loosely-captured exomoons around giant planets in habitable zones can survive to stabilize into long-lived orbits.
Table \ref{tab:gen} shows the fraction that stabilize for each mass/distance scenario.
The exomoons stabilized much easier in the F0 case than the M0 case, with the solar-mass case in-between.
The reason for this is readily apparent from the observation that while Hill radius scales linearly with distance from the star, the equilibrium temperature scales as the inverse square of distance.
Thus, the super-solar mass star allows a much wider Hill radius at a habitable distance than for a planet of equal mass around a sub-solar mass star.
This larger Hill radius provides the post-capture orbit much more room to move around, allowing longer period exomoons to stabilize.
The truncation of periods greater than two days for the M0 case in Figure \ref{fig:circ} is a clear result of this.

However, it also apparent from Figure \ref{fig:circ} that the fast stellarcentric orbits of the M0 case allow for much more rapid exomoon circularization that for the larger stars.
This clearly shows the effect of Kozai-pumped eccentricities accelerating tidal decay, as otherwise the timescales would be independent of the distance from the star.
The median circularization timescales for the M0 cases are of order $10^4$ years, $10^5$ years for the solar-mass star, and $10^6$ years for the F0 star.
All of these timescales are very short relative to the lifetime of the star, but $10^6$ years may be long enough that the exomoon's orbit could be externally perturbed by a planet or planetesimal disk.
On the other hand, as Figure \ref{fig:tale} shows, the semimajor axis decay typically happens at least an order of magnitude faster than full circularization.
Therefore, it is a reasonable assumption that nearly all these orbits would stabilize before any external perturbation would disrupt them.

To gauge the amount of tidal heating on the satellite, we estimated the amount of energy lost from the system from the difference in the initial and final orbits and spins.
For the orbits, we first found the energy difference between the initial and final states.
Then, we muliplied this an estimate of the fraction of energy that went into the satellite using the masses and tidal $Q$ of the objects, $Q_{planet}M_{sat}/(Q_{planet}M_{sat}+Q_{sat}M_{planet})$.
This term was usually near unity, as the planet was assumed to have a very large $Q$.
We then estimated the change in rotational energy of the satellite, subtracted this from the change in orbital energy, and divided the result by the circularization time to produce a heating rate.
The rate was dominated by the orbital term, as the initial rotation periods were not too different from the final orbital periods.
Generally, this rate was higher per unit mass for larger satellites (which circularized faster) and closer orbits.
For the Earth-mass/Jupiter-mass case, the median rates were 0.002 mW/kg around a F0 star, 0.02 mW/kg around a solar-mass star, and 0.5 mW/kg around a M0 star.
These rates are higher than for either long-lived \citep[$\approx10^{-8}$ mW/kg;][]{Desch2009} or short-lived chondritic radioisotopes \citep[$\approx10^{-4}$ mW/kg;][]{Schubert2007}.
A more through analysis is beyond the scope of this work, but these rates could create significant short-term heating on the exomoons.

Table \ref{tab:gen} also shows the fates of the exomoon orbits which did not survive to circularize.
The systems around the solar mass star were by far the most prone to separating beyond the Hill radius, with roughly equal impacts as separations.
The M0 cases, on the other hand, actually impacted the planet in three-quarters of the cases, due to the very small Hill radius not allowing the periapse to shrink very far before impacting.
A majority of F0 cases survived in all scenarios, with the large Hill radius allowing most orbits to circularize.
Therefore, for a given capture rate, potentially habitable captured exomoons should be the most common around stars larger than a solar mass.

Figure \ref{fig:inc} shows that the majority of the resulting orbits, especially around a Jupiter-mass planet, are nearly coplanar with the stellarcentric orbit.
This is partially due to the assumed low obliquity of the planet aligning its gravitational quadrapole with the direction of the star.
However, as shown by the high inclination of Triton (5.9 day period), an orbit does not need to be very large for this effect to be minimized.
And indeed, the orbits around Neptune-mass planets show a much larger diversity of inclinations for periods less than two days, especially for Earth-mass satellites.
Beyond this point, solar torques begin to dominate and force nearly all orbits coplanar.
Since longer period orbits produce larger TTVs, it is therefore likely that the first exomoons detected will be coplanar with the stellarcentric orbit, 
and it will take much more observations to detect short-period, high-inclination exomoons.

Also congruent with Triton, Figure \ref{fig:inc} shows that there is no real preference for either prograde or retrograde orbits.
The secular perturbations of Kozai cycles do not allow the directionality of the orbit to change in all but the most inclined initial cases.
Thus, each orbit preserves its original direction, evolving towards the plane of the stellarcentric orbit.
All of the final orbits are circular and have semimajor axes less than 28\% of the planet-moon system's Hill radius, 
which as summarized by \citet{Donnison2010}, means that they should be stable over the lifetime of the star.
Unfortunately, \citet{Kipping2009} shows that TTV and TDV are both degenerate for prograde/retrograde inclinations, and so measuring the relative fractions may be very difficult.
However, \citet{Simon2010} show that it may be possible to break this degeneracy with very precise measurements of the Rossiter-McLaughlin effect.
Since secular perturbations cannot reverse the direction of a satellite's orbit, detection of a retrograde satellite around an exoplanet would be confirmation that this capture process occurred.

The median TTV and TDV for each simulation set are given in Table \ref{tab:trans}.
The TTV do vary over a large range, but the upper end of that range is just within the limits of detection for modern transit systems. 
The Kepler mission has a regular cadence of approximately every 30 minutes, and can run at a cadence of up to once per minute \citep{Koch2010}.
This appears sufficient to detect an Earth-mass exomoon in most cases, and just enough to detect a Mars-mass exomoon in the best cases.
Larger stars offer wider orbits (and thus stronger TTVs), but much more infrequent transits. 
Solar mass to early K stars therefore seem to offer a good balance between TTV strength and transit frequency.

The Transit Duration Variation (TDV) introduced by \citet{Kipping2009} allows the unambiguous detection of an exomoon from transits alone.
For the systems described in that work, the TDV was of similar duration to the TTV.
However, we find that the lower planetary masses and much wider stellarcentric orbits of our study result in relatively weak TDVs for nearly all cases.
Thus, it is unlikely that a potentially habitable exomoon will be detected by TDV alone; TDV could, though, allow for follow-up transit observations to confirm the existence of a TTV-detected exomoon.

These results compare well with previous estimates, though both of those simulations assumed much wider orbits than were usually produced by the KCTF model.
The consequence of this is both reduce the TDV signal relative to that assumed by \citet{Kipping2009} and the photometric signal of \citet{Szabo2006}.
On the other hand, Table \ref{tab:trans} shows that even with these closer orbits, some exomoons are still within the range of detectability.
The combination of TTV and TDV may offer a stronger detection signal than photometry for these orbits, though both could detect some of the orbits produced.

%\begin{deluxetable*}{lllcccc}
%\tabletypesize{\scriptsize}
%\tablecaption{Median Exomoon orbital period, Transit Timing Variation (TTV), and Transit Duration Variation (TDV) for full-evolved exomoon systems.\label{tab:trans}}
%\tablewidth{0pt} \tablenum{2}
%\tablehead{\colhead{Star} & \colhead{Planet} & \colhead{Moon} & \colhead{Period (d)} & \colhead{TTV (min)} & \colhead{TDV} & \colhead{edge-on (min)}}
%\startdata
%\multirow{6}{*}{Sun} & \multirow{3}{*}{Jupiter} & Earth & 2.17 & 5.44 & 0.114\% & 0.93 \\
% &  & Mars & 2.34 & 0.59 & 0.012\% & 0.10 \\
% &  & Titan & 2.52 & 0.12 & 0.002\% & 0.02 \\
%\cline{2-7}
% & \multirow{3}{*}{Neptune} & Earth & 1.65 & 36.69 & 0.835\% & 6.41 \\
% &  & Mars & 1.89 & 4.11 & 0.089\% & 0.69 \\
% &  & Titan & 2.16 & 0.86 & 0.018\% & 0.14 \\
%\tableline
%\multirow{6}{*}{F0} & \multirow{3}{*}{Jupiter} & Earth & 3.68 & 11.33 & 0.111\% & 1.45 \\
% &  & Mars & 4.26 & 1.22 & 0.011\% & 0.15 \\
% &  & Titan & 4.14 & 0.26 & 0.002\% & 0.03 \\
%\cline{2-7}
% & \multirow{3}{*}{Neptune} & Earth & 2.38 & 75.70 & 0.860\% & 10.60 \\
% &  & Mars & 3.79 & 8.51 & 0.083\% & 1.03 \\
% &  & Titan & 3.50 & 1.78 & 0.018\% & 0.22 \\
%\tableline
%\multirow{6}{*}{M0} & \multirow{3}{*}{Jupiter} & Earth & 0.94 & 1.63 & 0.122\% & 0.54 \\
% &  & Mars & 1.00 & 0.18 & 0.013\% & 0.06 \\
% &  & Titan & 1.00 & 0.04 & 0.003\% & 0.01 \\
%\cline{2-7}
% & \multirow{3}{*}{Neptune} & Earth & 0.85 & 10.97 & 0.836\% & 3.36 \\
% &  & Mars & 0.79 & 1.22 & 0.098\% & 0.39 \\
% &  & Titan & 0.83 & 0.26 & 0.020\% & 0.08 \\
%\enddata
%\end{deluxetable*}

\section{Conclusions}

Several general conclusions can therefore be drawn from our simulations:
\begin{enumerate}
\item Loosely-captured exomoons around giant planets in habitable zones can survive to stabilize into long-lived orbits.
\item The timescale for them to stabilize is enhanced by Kozai torques, and is generally less than a few million years.
\item Most of the surviving orbits are close to the plane of the exoplanet's orbit, but show no prograde/retrograde preference.
\item Some of the resulting orbits should be currently detectable (especially for Earth-mass satellites), with the transit timing variation much stronger than the duration variation.
\item The most promising targets for detecting potentially habitable exomoons by TTV appear to be Neptune-mass exoplanets orbiting stars of solar mass or slightly below.
\end{enumerate}
\vspace{5 mm}
\acknowledgments
Special thanks to Travis Barman for donating the approximately 5000 CPU hours it took to complete this project.

\bibliographystyle{apj}

\end{document}